\documentclass[12pt,preprint]{aastex}

\shorttitle{TRIS I} \shortauthors{Zannoni et al.}

\begin{document}

\title{TRIS I : \\ Absolute Measurements of the Sky Brightness Temperature
 \\
    at 0.6, 0.82 and 2.5 GHz }

\author{M. Zannoni\altaffilmark{1}, A. Tartari, M. Gervasi\altaffilmark{1,2},
 G. Boella\altaffilmark{2}, G. Sironi\altaffilmark{1},\\ A. De Lucia and A.
Passerini} \affil{Physics Department, University of Milano
Bicocca, P.zza della Scienza 3, I20126 Milano Italy}

\author{F. Cavaliere}
\affil{Physics Department, University of Milano, via Celoria 16,
I20133 Milano Italy} \email{mario.zannoni@mib.infn.it}

\altaffiltext{1}{also Italian National Institute for Astrophysics,
INAF, Milano.} \altaffiltext{2}{also Italian National Institute
for Nuclear Physics, INFN, Milano-Bicocca.}

\begin{abstract}
At frequencies close to 1 GHz the sky diffuse radiation is a
superposition of radiation of Galactic origin, the 3 K Relic or
Cosmic Microwave Background Radiation, and the signal produced by
unresolved extragalactic sources. Because of their different
origin and space distribution the relative importance of the three
components varies with frequency and depends on the direction of
observation. With the aim of disentangling the components we built
TRIS, a system of three radiometers, and studied the temperature
of the sky at $\nu =0.6$, $\nu = 0.82$ and $\nu = 2.5$ GHz using
geometrically scaled antennas with identical beams (HPBW =
$18^{\circ} \times 23^{\circ}$). Observations included drift scans
along a circle at constant declination $\delta=+42^{\circ}$ which
provided the dependence of the sky signal on the Right Ascension,
and absolute measurement of the sky temperature at selected points
along the same scan circle. TRIS was installed at Campo Imperatore
(lat. = $42^{\circ}~26'$ N, long.= $13^{\circ}~33'$, elevation =
2000 m a.s.l.) in Central Italy, close to the Gran Sasso
Laboratory.
\end{abstract}


\section{Introduction} \label{intro}
The diffuse radiation from the sky is a superposition of
components. At frequencies between few tens of MHz and few tens of
GHz, in terms of brightness temperature we can write
\begin{equation}
T_{sky}(\alpha,\delta,\nu)=T_{gal}(\alpha,\delta,
\nu)+T_{CMB}(\nu)+T_{uers}(\nu) \label{eq:Tskygen}
\end{equation}
\par\noindent where $\alpha$ and $\delta$ are the Right Ascension and
declination of the point at which the telescope axis is aimed.
$T_{gal}$ is the Galactic contribution: partially polarized,
anisotropically distributed, it tracks the Galactic structure. Its
frequency spectrum is a power law
\begin{equation}
T_{gal}(\nu, \alpha, \delta)  = K(\alpha, \delta) ~\nu
^{-\beta({\alpha,\delta})} \label{eq:Tgal}
\end{equation}
\noindent with spectral index $\beta$ ranging from 2.1 to 3.4
depending on the relative weight of thermal and synchrotron
emission, on the energy spectrum of the Cosmic Ray Electrons and
on the Galactic Magnetic Field. Compared to the other components
of the diffuse radiation it is a foreground and dominates the sky
at frequencies below $\sim 1$ GHz.

\noindent $T_{CMB}$ is the Cosmic Microwave Background. Relic of
the Big Bang, it is substantially unpolarized and isotropically
distributed. Its flat frequency spectrum is consistent with the
emission of a blackbody with a temperature of $(2.725 \pm 0.001)$
K \cite{Fixsen2002}. Compared to the other components it is a true
background. At $\nu \geq 1$ GHz $T_{CMB}$ definitely overcomes
$T_{gal}$ and dominates the sky up to $\nu \sim 100$ GHz, above
which thermal emissions from an irregular distribution
of dust with physical temperature $\geq 20$ K, gradually
overwhelms the other components.

\noindent $T_{uers}$ is a blend of unresolved extragalactic radio
sources isotropically distributed. Its frequency spectrum is a
power law
\begin{equation}
T_{uers} = K_{uers} \nu^{-\gamma_{uers}}  \label{eq:Tex}
\end{equation}
with $\gamma_{uers} \sim 2.70$ \cite{Gervasi2008b}. \newline
\noindent Each of the above components carries important
astrophysical and/or cosmological information, particularly at
decimetric wavelengths. At these frequencies for instance we can
expect a dip in the flat spectrum of $T_{CMB}$, whose detection
can provide direct information on $\Omega_b$ \cite{Burigana1991},
the Universe baryon density. In this same region the slope of the
power law frequency spectrum of $T_{gal}$ changes, revealing
perhaps a knee in the energy spectrum of the cosmic ray electrons
responsible for the galactic synchrotron radiation. To disentangle
these components, coordinated, multifrequency observations of
extended areas of sky are necessary. They will form a data base
from which the desired information can be extracted modelling the
components and optimizing the model parameters. At frequencies
close to 1 GHz it is possible to find in literature maps of
$T_{sky}$ which cover all the sky or large parts of it (see Table
\ref{tab1}). Their accuracy is however insufficient to answer many
of the questions put forth by present day astrophysical and
cosmological models. These maps have been in fact obtained
combining data collected at different sites and/or at different
times and frequently show artificial structures, e.g. {\it
stripes}. Once {\it destriped} (see for instance
\cite{Platania2003}), these maps can be used to extract the
spectral index of the galactic signal $T_{gal}$. The accuracy of
the zero level of the absolute scale of temperature of the same
maps is however still insufficient to disentangle the radiation
components with the accuracy today required by cosmology. To
overcome this difficulty we made absolute and differential
measurements of the diffuse radiation at three frequencies, using
a set of three radiometers (therefore the project is said TRIS),
at 0.6, 0.82 and 2.5 GHz. In this paper we present results of
these observations. In \cite{Gervasi2008a} and \cite{Tartari2008}
(from now on cited as Paper II and III), the astrophysical and
cosmological implications of TRIS data are discussed.

\section{Absolute measurements of temperature}
\label{Accabs} Measuring the absolute value of the {\it brightness
temperature}, $T_{sky}$, of a patch of diffuse radiation is a
conceptually simple but not easy task. Large systematic effects
are in fact expected.

A radiometer aimed at the sky gives the {\it antenna temperature},
$T_a$, from which we obtain
\begin{equation}
T_{sky} = \frac{T_a - T_{atm}- T_{env}}{1 - (T_{atm}/T^{0}_{atm})}
\label{eq:Target}
\end{equation}
\par\noindent where $T_{atm}$ and $T_{env}$ are the noise
temperatures of the signals from the atmosphere and the
environment above and around the radiometer and $T^{0}_{atm}\simeq
240 K$ is the physical temperature of the atmosphere
\cite{US_STANDARD_ATMOSPHERE_1976}. References for calculations
are presented in the following Section \ref{skytemp}.
\newline \noindent Since the diffuse radiation fills the antenna
beam, the absolute value of $T_a$ is
\begin{equation}
T_a = T_{cold}^{eff} + (S_{sky} - S_{cold}) G
\label{eq:Tantgen}
\end{equation}
\noindent where
\begin{equation}
G = \frac{T_{warm}^{eff}-T_{cold}^{eff}}{S_{warm}-S_{cold}}
\label{eq:gain}
\end{equation}
\par\noindent is the system {\it gain} or {\it conversion factor}
and $S_{sky}$, $S_{cold}$ and $S_{warm}$ are the radiometer
outputs produced by the sky target and by two known sources ({\it
calibrators}), a {\it cold} and a {\it warm} load. $T_x^{eff}$,
the {\it effective temperature} of calibrator $x$, is the
convolution of the calibrator brightness temperature and the
antenna beam pattern. Effects of antenna and receiver properties
(gain, bandwidth, physical temperature and attenuation of the
components between the antenna mouth and the radiometer output)
are included in $G$.

With one important exception \cite{Stankevich1970} astronomical
objects have been rarely used as absolute calibrators. In fact
unless the source is isolated, it completely fills the antenna
beam and its brightness distribution is precisely known, the
accuracies of $T^{eff}_x$, $G$ and $T_a$ are poor.  Artificial
blackbodies, shaped to fit the antenna aperture, to match the
antenna impedance and to fill the antenna beam, are more
convenient calibrators, because their effective and brightness
temperatures coincide. They have been used \cite{Bensadoun1992} at
frequencies $\nu > 1$ GHz ($\lambda < 30~cm$) where the radiation
wavelength, the antenna aperture $A_e = \lambda^2/\Omega_a$
($\Omega_a$ is the antenna solid angle), and the blackbody
dimensions are reasonably small. Small dimensions make also
possible to bring the radiometer in space (e.g. \cite{Mather1994})
or at balloon altitudes (e.g. \cite{Kogut2006}) where atmospheric
and environmental effects are negligible or absent and $T_a\simeq
T_{sky}$ (see \ref{skytemp} and Equation (\ref{eq:Tsky})).

\noindent Below 1 GHz however the antenna dimensions and the
wavelength are large and discourage the construction of artificial
sources which fit the antenna mouth and fill the antenna beam.
Moreover, at these low frequencies observations from space or at
balloon altitudes are extremely difficult and calibrations are
better made injecting at some point between antenna and receiver,
through a multi throw switch, the noise produced by very compact
line terminations (the so called {\it dummy loads}) properly
cooled or warmed.
\newline \noindent
In this case the effective temperature of the calibrator is:
\begin{equation}
 T_{l}^{eff} = \Bigl[T_{l}^0 e^{-\tau_{c}} + \int_{0}^{L} T_{c}^{0}(x)e^{-\tau(x)}({\frac{d\tau}{dx}})dx \Bigr] (1-r^2) + r^2 T_{RX}^{eff}
 \label{eq:Teff}
\end{equation}
where $T_{l}^{0}$ and $T_{c}^{0}(x)$ are the physical temperatures
of the load and the transmission line {\em connecting} the load to
the switch, $x$ is a coordinate running on the line,
$\frac{d\tau}{dx}$ is the specific attenuation (neper per unit
length) of the line, $e^{-\tau_{c}}$ is the transmission of the
line, $r^2$ the power reflection coefficient of the load seen
through the line and $T_{RX}^{eff}$ is the noise temperature
radiated by the receiver (see \ref{receiver}).

Usually the accuracy of $G$ is good but $T_a$, being measured at
the switch positions, must be corrected (see for instance Equation
(\ref{eq:Tant})) for the transmission $e^{-\tau_{lc}}$ and {\it
thermal noise}
\begin{equation}
T_{noise} = \int_0^{\tau_{lc}}{T_{0}(x)e^{-\tau(x)} d\tau} =
\overline{T}_{0}(1-e^{-\tau_{lc}}) \label{eq:Tnoise}
\end{equation}
of the lossy components (LC) at temperature $T_{0}$ distributed
along the line between the switch and the antenna mouth. To limit
the statistical and systematic uncertainties on $T_{noise}$ the
temperature of these components must be low and stable. In fact
Equation \ref{eq:Tnoise} gives:
\begin{equation}
(\delta T_{noise})^2 ~\simeq (T_o ~{\delta
\tau_{lc}})^2+(\tau_{lc}~\delta T_0)^2~~~~(\tau_{lc} \ll 1)
\label{eq:DeltaTnoise}
\end{equation}
e.g. $0.3 \leq \delta T_{noise} \leq 3$~K if $T_0 \simeq 300 K$
and $10^{-3}\leq \tau_{lc} \leq 10^{-2}$. For the above reasons
the values of $T_{sky}$ measured at frequencies close and below 1
GHz have been so far less accurate than the values measured at
higher frequencies.

Reviews in literature (see for instance \cite{Sironi1999},
\cite{Zannoni1999}, \\ \cite{Salvaterra2002} and Table \ref{tab2})
show that below 10 GHz the accuracy of $T_{sky}$ rapidly decreases
with frequency and it is very poor below 1 GHz. Moreover between 1
and 0.5 GHz the large dispersion of the error bars assigned by
different observers to their results suggest that sometimes in the
past not all the systematic effects were recognized. The frequency
dependence of the accuracy of $T_{sky}$ is especially evident in
coordinated experiments  like the multifrequency observations made
by the White Mt. collaboration \cite{Smoot1985}, the South Pole
collaboration \cite{Bersanelli1993} and by Arcade
\cite{Kogut2004}. These experiments cover the frequency interval
$2 - 100$ GHz and approach the frequency region close to 1 GHz
where neither $T_{CMB}$ nor $T_{gal}$ are dominant and the models
of $T_{gal}$ and $T_{uers}$ are insufficient to extract with the
desired accuracy the cosmological signal. Additional observations
at even lower frequencies were subsequently made at 1.4 GHz, 0.82
and 0.61 GHz (\cite{Bensadoun1993}, \cite{Sironi1990} and
\cite{Sironi1991}). Unfortunately those last observations used
different systems and were not coordinated among them nor with
previous observations at the same or at higher frequencies. TRIS
has been planned to carry on coordinated observations between 0.6
and 2.5 GHz with the aim of improving the accuracy of the
existing data.

\section{TRIS} \label{tris}
\par TRIS is a set of three absolute radiometers which operate
in total power configuration. All the radiometers have the same
antenna beam, similar receivers and include an External Calibrator
unit and an Internal Calibrator unit (see Figure \ref{F1:block}).

The frequencies of operation (0.6 GHz ($\lambda$= 50 cm), 0.82
GHz ($\lambda$= 36 cm) and 2.5 GHz ($\lambda$= 12 cm)) were chosen
because they span a frequency region where the ratio $(T_{gal} /
T_{CMB})$ is $>1$ at low frequencies and $<1$ at high frequencies
and because were used in the past for similar observations.

\subsection{Antennas} \label{antennas}
\par\noindent To have the same beam the three antennas are
geometrically scaled (i.e. their linear dimensions are the same in
wavelength units). We chose pyramidal rectangular horns with
corrugated (thin corrugations) E-plane walls and a corrugated
E-plane corona at the mouth (see Table \ref{tab3}) because they
have measurable and/or calculable electromagnetic properties.
These antennas have a beam solid angle $\Omega_a$ wide enough to
dilute the contribution of isolated sources well below the signal
produced by the diffuse radiation but capable to keep track of the
main features of the radiation spatial distribution. Side and back
lobes are low ($\leq -30$ dB, Figure \ref{F2:beam}) to minimize
spill-over from the ground and interferences. The e.m. signal is
extracted from the antenna through a tunable (five tuning stubs)
coaxial - waveguide transition. A prototype of the complete
antenna at 8.2 GHz was built and tested in anechoic chamber. Then
the 2.5 GHz was made scaling the 8.2 GHz unit and optimizing the
waveguide - coaxial transition. The 0.6 and 0.82 GHz horns were
then scaled from it.
\newline \noindent Horn walls, corrugations, corona, waveguide
sections and waveguide-coax transition are made of anticorodal
riveted plates. These parts are fastened with screws for easy
disassembling and transportation (before going to Campo Imperatore
the 0.82 and 2.5 GHz horns were used at the Amundsen Scott base at
South Pole). Aluminum tape on all the joints provides optimum
electrical contacts among the various parts and makes them
insensitive to variable weather condition and oxidation.

The 0.6, 0.82 and 2.5 GHz units were too large for the available
anechoic chambers. The horn beam was therefore measured on the 8.2
GHz prototype in anechoic chamber and at a test range, putting
special care on side and back lobes. Results, presented in Figure
\ref{F2:beam}, give a beam solid angle of $18^{\circ}(E) \times
23^{\circ}(H)$ (HPBW). Tests on the 0.6, 0.82 and 2.5 GHz units,
made looking at the transit of the sun, confirmed it.

The transmission $e^{-\tau}$ of the components which bring the
signals from the antenna throat to the receiver were measured at
0.6, 0.82 and 2.5 GHz with a Vector Network Analyzer (Agilent
8510C) in laboratory. The attenuations of the horn between mouth
and throat was calculated. Results are shown in Table \ref{tab4}.

At Campo Imperatore the three antennas were on an almost East West
row, on stands which allow to steer the beam along the E-plane.
This plane is titled $7^\circ$ from the N-S plane in direction
N$\rightarrow$NE and S$\rightarrow$SW. This plane crosses the
southern profile of the surrounding mountains at azimuth
187$^\circ$ where there is a brief levelling (see Figure
\ref{F3:horiz}).

\subsection{External Calibrator}
A SPTT (Single Pole Triple Throw) switch is used to link the
receiver to the antenna, the cold calibrator or the warm
calibrator. The switch is a source of thermal noise. The maximum
accuracy one can reach in evaluating it, is set by the
repeatability of the switch attenuation and by the stability of
its temperature (see eqs.(\ref{eq:Tnoise}) and
(\ref{eq:DeltaTnoise})). To minimize this noise and its
fluctuations, the switch is cooled at temperatures comparable to
the sky temperature and is stabilized putting it in a bath of
liquid helium. Because no commercially available units satisfied
our needs, we studied and built a cryogenic, purely passive,
manually driven, resonant coaxial SPTT switch based on $\lambda/4$
lines \cite{Zannoni1999}. With $\tau \leq 0.01 dB$ and $\delta
\tau \leq 2.3\times10^{-3}$ it gives $\delta T_{noise, switch}
\leq 10$ mK when cooled at 3.967 K, the boiling temperature of
liquid helium at Campo Imperatore (2000 m a.s.l.). When cooled the
switch can be driven through fiber glass rods which come out from
the dewar housing it.

\subsubsection{Cryogenic Waveguide}\label{cryo}
To minimize their thermal noise all the lossy components
distributed between the horn throat and the receiver input are
cooled. The cryogenic SPTT switch,  the waveguide-coaxial
transition and two 50 $\Omega$ dummy loads are mounted at the bottom of
a dewar (see Figure \ref{F4:dewar}) where they are always plunged
in liquid helium (LHe).  The third load is at the top of the
dewar, at a stable temperature close to the external ambient
temperature. A special stainless steel waveguide section brings
the electromagnetic signal from the waveguide transition to the
dewar top where it can be attached to the horn throat.

The stainless steel guide section is a thermal insulator. A window
made of a polyethylene sheet in the collar at the top of the
stainless steel section and two fluorglass (a special fabric
transparent to radio and microwaves, opaque to IR) sheets in the
stainless steel waveguide limit the input of thermal radiation. Holes
in the fluorglass windows allow a uniform distribution of the cold
helium vapors inside the waveguide.

A cable which goes through the flange at the dewar mouth links one
of the two cold dummy loads to the circulator third port in the
Internal Calibration Unit (see Figure \ref{F1:block}). A second
cable links the switch output to the receiver input. The three
inputs of the switch go to: 1) the antenna coaxial input through a
very short piece of cable at LHe temperature; 2) the second cold
dummy load (the {\it cold source}); 3) the load at the top of the
dewar (the {\it warm load}) through a long cable. Sensors monitor
the level of the liquid helium and the temperatures of switch,
cold and warm loads, waveguide walls and cables inside and outside
the dewar.

A 250 l dewar can house the 0.6 or the 0.82 GHz system. It is
sufficient to carry on observations for more than 48 hours without
refilling. The 2.5 GHz system is in a 100 l dewar. Once filled
with helium it allows continuous observation for about 90 hours.

The effective temperatures of the cold loads are larger than 3.967
K, the boiling temperature of LHe at the observing site, because
of the unavoidable e.m. mismatches which carry in a fraction of
the noise temperature radiated by the receiver (see Equation
\ref{eq:Teff} and Table \ref{tab5}). Their  values however can be
calculated. The high value of $T_{cl}^{eff}$ at 2.5 GHz is a
consequence of the increasing difficulties of matching resonant
systems when the wavelength decreases. The complete structure of
the horn and Cryogenic Waveguide is shown in Figure
\ref{F4:dewar}.

\subsubsection{Warm Waveguide}\label{warm}

A more simple calibrator based on a small dewar containing just
the SPTT switch and the three dummy loads was also prepared. It
requires a limited quantity of LHe. It was  used for preliminary
tests and for 0.82 GHz absolute measurements when the dewar of the
cryogenic waveguide failed. The final accuracy one gets on
$T_{sky}$ is however worse compared to the accuracy one gets using
the Cryogenic Waveguide because of the noise produced by the lossy
components, including the waveguide - coaxial transition, at
ambient temperature.

\subsection{Internal Calibrator and Reflectometer}\label{reflect}
\par A combination of circulators, directional couplers, solid
state noise source, matched dummy loads and electrically driven
switches form a unit, mounted between the receiver input and the
Low Noise Amplifier (see Fig.\ref{F1:block} IntCal block). It can be used to:

\noindent i) check the e.m. matching of the components attached at
the receiver input. This is done turning on the solid state noise
source and sending its signal toward the receiver input. The
measured variations of the receiver outputs $\Delta O_{inj}$ and
$\Delta O_{ref}$  when the receiver input is, respectively, short
circuited or connected to the device under test, give the {\it
power reflection coefficient} $r^2 = {\Delta O_{ref}/\Delta
O_{inj}}$ ({\it reflectometer mode}).

\noindent ii) check the receiver gain sending the noise generator
signal into the receiver ({\it internal calibration mode}).

\noindent iii) set to a known level the effective temperature
$T_{RX}^{eff}$ of the noise the receiver radiates toward the
antenna. This is the {\it effective noise temperature}
$T_{N}^{eff}$ of the load at the third port of the circulator (see
Equation (\ref{eq:Teff}) and Figure \ref{F1:block}). When this
load is cooled at liquid helium temperature (see below)
$T_{RX}^{eff}$ can be of few tens of K.

The internal calibrator and the reflectometer are used for regular
checks of the radiometer performance during the observations.
Cross checks are made on site with a scalar network analyzer.

\subsection{Receivers} \label{receiver}
TRIS receivers are standard heterodyne chains with double
frequency conversion, total gain of $\sim 100$ dB, used in total
power configuration (see Figure \ref{F1:block} and Table
\ref{tab5}). The frequency of the Local Oscillator LO1 can be
adjusted digitally (see Figure \ref{F1:block}) so one can tune the
observing channel among 256 adjacent frequencies separated by
$\Delta\nu$, symmetrically distributed around the system nominal
frequency $\nu_o$. $\Delta\nu$ is 75 KHz at 0.6 and 0.82 GHz,
spanning a range of 20 MHz, and 750 KHz at 2.5 GHz, over a range
of 200 MHz. The integration bandwidth is 0.3 MHz at 0.60 and 0.82
GHz and 3 MHz at 2.5 GHz one.

The post detection integration time constant is 10 sec. After
integration the receiver output is sampled and digitally converted
every 4 sec. Each sampled value, the housekeeping data and the
time signal are written in a record and stored on the hard disk of
the Personal Computer which drives the local oscillator LO1. The
time signals arrive from a master clock locked to radio signals
regularly transmitted by I.E.N. Galileo Ferraris in Turin. The
same clock drives the computer clock and guides the observational
sequence. To avoid digital noise the 20-bit AD converter is
integrated into the detector channel and data are transmitted via
optocoupler to the data recording system.

\par Measurements in total power configuration can be affected
by gain instabilities. To compensate for them the receiver
detection channel includes zero-bias Schottky diodes and
temperature stabilization of the DC section (active control with
accuracy better than $0.1~^{\circ}C$). The receiver temperature is
set at stable values ( $\pm 0.1~^{\circ}C$) chosen between
$25~^{\circ}C$ and $35~^{\circ}C$ for excursions of the external
temperature from well below zero up to the preset value (usually
$+35~^{\circ}C$). Gain variations can be monitored and, when
necessary, recovered looking at the amplitude of the internal
calibrator signal (see below). They are usually small ($\leq 2 \%
$) except in rare occasions during observations around the local
noon when clear sky and no breeze may bring the external
temperature of the receiver box above the preset value. When this
situation occurs data are rejected.

\section{Observations}
Between 1990 and 1993 prototypes of the three radiometers and
antennas were installed at Campo Imperatore (lat. =$42^o~ 26'$ N,
long.= $13^o~ 33'$ E), a plateau 2000 m a.s.l.. The site is
reasonably shielded against unwanted signals from the horizon by a
circle of low elevation mountains (see Figure \ref{F3:horiz}).
Served by road in summer and by cable car in winter, the site is
on the vertical of the underground Laboratori Nazionali del Gran
Sasso (LNGS) which provided logistical support. This accommodation
is a reasonable compromise between ideal places not easily
accessible (like Antarctica)  or isolated places with no
facilities available (like White Mt. (California) or Alpe Gera
(Italian Alps)) we used in the past.

Various tests were made at Campo Imperatore between 1991 and 1995.
They showed that in spite of the mountain circle, interferences
from horizon directions were occasionally present and forced us to
observe only at the zenith. The final installation of TRIS
radiometers was made between summer 1997 and summer 1998. After
repeated observations dedicated to find clean frequency channels,
differential measurements (drift scans) began. They continued in
1999 when tests of the cryogenic front ends were completed.
Absolute measurements of the sky temperature were made in June and
October 2000. Drift scans were then resumed and continued until
May 2001 when antennas were removed from Campo Imperatore.

\subsection{Modes of operation} \label{modes}
\par Three modes of operation have been used with TRIS:

\par{\it Interference Search Mode}. The PC drives the Local Oscillator
and tunes cyclically the receiver  through the 256 frequencies
distributed around the receiver nominal frequency $\nu_o$ (10
minutes/channel). Simple statistics (mean and standard deviation)
have been used to recognize channels disturbed by interferences
and to choose quiet channels where observations were possible.
  Searches lasting at least 24 h were repeated at various epochs.
At 0.6 and 0.82 GHz we found frequencies free from interferences.
The best were 0.6005 and 0.8178 GHz. At 2.5 GHz no frequency
completely free from interferences was found. The best one was
2.4278 GHz. These frequencies are those used for the final
observations in both the following modes.

\par {\it Absolute Measurement Mode}. The antenna is aimed at the
zenith. The horn throat is attached to the collar of the cryogenic
front end. A couple of hour after filling (or partial refilling)
with LHe, the system is in thermal equilibrium and observations
are possible. Moving the switch manually the receiver input is
cyclically connected to the Cold Source (receiver output
$S_{cold}$), the Antenna (receiver output $S_{sky}$) and the Warm
Source (receiver output $S_{warm}$). Each step of the cycle lasts
10 minutes during which 15 records/minute are stored on the PC
hard disk. Each record contains $S_x$, UT time, Julian day,
temperature of loads and lossy components between switch and
antenna mouth, housekeeping data. The cycle of three steps is
repeated few times then the switch is set on the Antenna position
and observation goes on for about one hour recording $S_{sky}$. At
regular time intervals the solid state noise source is set on and
data necessary to check gain stability and power reflection
coefficients $r^2$ of the source at the receiver input are
recorded. \newline \noindent Absolute measurements are made only
at nighttime, when the observing conditions are good and the dewar
full of LHe well above the top of the waveguide coaxial
transition. At 0.6 GHz observations with the Cryogenic Waveguide
went on regularly. At 0.82 GHz we began observation in June using
the Cryogenic Front End but the dewar failed because of a leakage
in the vacuum tank. We fixed it but in October, when after the
0.6 GHz run we moved at 0.82 GHz, the dewar failed again and we
had no possibility to fix it again before leaving Campo
Imperatore. So at 0.82 GHz we have only absolute measurements made
as preliminary tests in June, using the cryogenic switch cooled
with liquid helium and the completely warm antenna. This system
worked very efficiently, but because all the horn components,
including the waveguide-coaxial transition, were at ambient
temperature, final results at 0.82 GHz are less accurate than at
0.6 GHz.\newline \noindent The absolute measurements at 2.5 GHz
with the Cryogenic Waveguide went on regularly but were plagued by
interferences.

\par {\it Drift Scan Mode}. Records with $S_{sky}$ and the
associated information are stored continuously every 4 sec while
the sky transits through the antenna beam aimed at the zenith. In
this mode we look at the variations of $S_{sky}$ with Right
Ascension and do not care of the signal zero level, provided it is
stable. The antenna can be attached to the dewar (no matter if
cold or warm) with the switch on position 2 or attached directly
to the receiver via an ambient temperature coaxial-waveguide
transition. Gain stability and antenna power reflection
coefficient are measured automatically two times per hour. In this mode
the radiometers can work unattended for weeks. Because data
collected at daytime and more noticeably around noon are usually
contaminated by the sun emission, to cover the entire
$(0^h~-~24^h)$ Right Ascension interval drift scans are repeated
months apart. \newline \noindent Drift scan data were collected
during the entire lifetime of TRIS. The more systematic
observations were made from 1998 on. At 0.6 and 0.82 GHz they are
sufficient to cover the entire 24 h circle of Right Ascension at
$\delta = +42^{\circ}$. At 2.5 GHz the great majority of the
observations were disturbed by interferences and no complete drift
scans were obtained.

\subsection{Data Reduction} \label{datared}

First of all the raw data collected in Absolute or Drift Scan mode
are {\em edited}. We reject all the records containing data
obtained:
\par i) at daytime (between half an hour before sunrise and half an hour
after sunset);
\par ii) less than six hour after the receiver was turned on;
\par\noindent or when
\par iii) interferences were evident on the records;
\par iv) the log book shows that the weather conditions were bad (rain,
snow);
\par v) the antenna reflection coefficients $r^2$ monitored by the receiver internal
reflectometer worsened (e.g because of water vapor condensation on
the antenna components at sunrise or sunset);
\par vi) the system gain or the noise level change or fluctuate by more than
$5 \%$.

The remaining  data are then divided in blocks, continuous time
series of homogeneous records, lasting at least 10 minutes. A
block starts when the switch is set in a position (or when drift
scan mode starts) and stops when the switch is moved to a new
position or whenever a major interruption of the time series
occurs.
 We call them {\it Absolute Sky, Cold, Warm} and {\it Drift Scan} or
{\it Differential} blocks (or records, or data), depending on the
observing mode and/or the switch position. Typical block lengths
are 5-20 minutes for absolute data. The blocks of differential
data can last hours.

Then we remove all the data collected in the first minute of a
block time series (to account for the response time of the system)
and continuous time series of differential data lasting less than
three hours and separated from the following homogeneous block of
data used for analysis by more than one hour.

Finally for each record we calculate and add to it $(\alpha,
\delta)$ of the antenna beam axis calculated from the recorded
values of time and Julian day, the effective noise temperatures
radiated by the receiver ~$T_{RX}^{eff}$ {\footnote{When loads are
well matched in Eq. (\ref{eq:Teff}) we can set $r^2\simeq 0$, get
the effective temperature of the circulator third port
$T_{N}^{eff}$ and set it equal to ~$T_{RX}^{eff}$. Then we repeat
the calculation of the effective noise temperature of the
circulator third port  as seen at the switch and use this value as
final $T_{RX}^{eff}$} and the effective noise temperatures of Cold
Load, Warm Load and Circulator third Port, respectively,
$T_{cold}^{eff}$, $T_{warm}^{eff}$ and $T_{N}^{eff}$ using
(Equation (\ref{eq:Teff})).

The resulting blocks are the data ready to be analyzed. They are
treated in different ways, depending on their type.

\subsection{Absolute values of sky temperature} \label{absmeas}

From pairs of time adjacent blocks of {\it cold} and {\it warm
data} the gain $G$ is obtained using Equation (\ref{eq:gain}) and
the average  values of $S_{cold}, T^{eff}_{cold}, S_{warm}$ and
$T^{eff}_{warm}$ on the block. Typical values are shown in Table
\ref{tab5}.

Then from each value of $S_{sky}$ in a {\it sky} block and $G$
from the nearest pair of {\it cold} and {\it warm} data we get the
instantaneous value of the signal temperature which arrives at the
switch:
\begin{equation}
T_{a,sw} = T^{eff}_{cold}+(S_{sky}-S_{cold}) G
\label{eq:Tainst}
\end{equation}

The system linearity, i.e. the stability of G and its independence
on the signal level, have been checked during all the calibration
campaigns, by looking at the amplitude of the Calibration Mark
(CM) when injected over very different level of signals: Cold Load
(He), Sky and Warm Load. It is worth to underline that the CM
provides an equivalent temperature in the same range of the
maximum variation of antenna temperature measured by our
experiment (about 20 K, 14 K and 1 K at 0.6, 0.82 and 2.5 GHz
respectively). CM amplitude was constant within statistical
fluctuations, setting a $0.4 \%$ upper limit to the deviation from
system linearity.

The antenna temperature at the horn mouth is then
\begin{equation}
T_{a}(\alpha,\delta) = \frac{1}{1 - r^2}\Biggl [\Biggl(
{\frac{T_{a,sw}-\int_0^{\tau_{tl}}{T_{tl}(x)e^{-\tau_{tl}(x)}d\tau}}
{e^{-\tau_{tl}}e^{-\tau_h}}}-\frac{\int_{0}^{\tau_{h}}
{T_{h}(x)e^{-\tau_{h}(x)}d\tau}}{e^{-\tau_{h}}}\Biggr) ~-~ r^2
T_{RX}^{eff}~\Biggr ] \label{eq:Tant}
\end{equation}

Here $T_{tl}$, $\tau_{tl}$, $T_h$ and $\tau_h$ are the
temperatures and the optical thicknesses of the lossy components
along the transmission line ($tl$) and the horn sections ($h$)
between the switch and the antenna aperture.

\subsubsection{\it Sky Temperature} \label{skytemp}

The sky brightness temperature is then:
\begin{equation}
T_{sky}(\alpha,\delta,\nu) = \Bigl[
{\frac{T_{a}(\alpha,\delta,\nu)-T_{RX}r^2}{1 -
r^2}}-T_{atm}^{0}(1-e^{-\tau_{atm}})-T_{env}\Bigr]/e^{-\tau_{atm}}
\label{eq:Tsky}
\end{equation}

\noindent where $\tau_{atm}$ and $T_{atm}^{0}$ are the optical
thickness and the average physical temperature of the atmosphere
above the antenna, $T_{env}$ the noise temperature of the
environment which reaches the antenna, usually through side and
back lobes, $T_{RX}^{eff}$ the temperature of the noise radiated
by the radiometer and $r^2$ the power reflection coefficient of
the antenna.

At nighttime, when the Sun contamination is absent, $T_{env} = T_{ground}+T_{inter}$ where $T_{ground}$,
the effect of the ground thermal emission, is the convolution of
$P_n(\theta,\phi)$, the normalized, three dimensional, beam
profile (see Figure \ref{F2:beam}) and a blackbody at ambient
temperature $T_{0}$, which fills the antenna beam up to $h(\phi)$,
the Campo Imperatore horizon profile, (see Figure \ref{F3:horiz}):
\begin{equation}
T_{ground} =
T_{0}~\int_{0}^{2\pi}d\phi\int_{0}^{h(\phi)}P_n(\theta,\phi)\sin
(\theta) d\theta  \label{eq:Tgr}
\end{equation}

\noindent $T_{inter}$ is the noise produced by radio
interferences. At 0.6 and 0.82 GHz channels free from
interferences were found (0.6005 GHz and 0.8178 GHz). Here the
interferences, if present, were completely buried in the system
noise. At 2.5 GHz also in the best channel at 2.4278 GHz, even when no
sudden changes of level were appreciable, the signal was unstable
and the noise anomalously high, probably because of a blend of
undesired signal from artificial sources at the horizon or below
it. This contribution has been measured surrounding the antenna
with ground shields which went well above the horizon visible from
the antenna mouth and reflected all the side and back lobes on the
sky.

For the atmospheric contributions we use values of absorption
calculated for various sites, including Campo Imperatore, by
\cite{Ajello1995} from a collection of vertical profiles of the
earth atmosphere collected for one year with sounding balloons
launched daily at meteo stations. Results are shown in Table
\ref{tab6}.

 The above values of $T_{sky}$, $T_a$,
$T_{a,sw}$ are oversampled because the TRIS sampling time (4 sec)
is shorter than the system time constant (10 sec). To make them
statistically independent  in each {\it sky block} data are binned
in groups of at least 50 sec and averaged. The average values and
standard deviations of $T_{sky}$ in a bin are considered
representative of the absolute brightness temperature of the sky
at ($<\alpha>$, $<\delta>$) only if they come from a block close
{\em (few tens of minutes)} to the pair of {\it cold} and {\it
warm} data used to get $G$ and $T^{eff}_{cold}$. This because $G$
and $T^{eff}_{cold}$ are helium level dependant and the level is
practically constant within this period of time.

\subsection{Drift scans of the sky temperature} \label{driftscan}
The edited time series of {\it drift scans} data are from
observations made in separate years at different times of the
year, in different conditions.
\par\noindent
During the drift-scan mode observations, every component of the
radiometers was at ambient temperature, with the receivers always
operating in a temperature controlled box. In this configuration,
the data were taken continuously for months. We decided to reject
daytime data. Moreover, bad weather conditions and interferences
didn't allow us to exploit all the night-time data collected,
therefore it took more than one year to cover all the
$\delta=+42^{\circ}$ sky circle since the last TRIS configuration
(antennas in vertical position, receivers in total power
configuration) was established. We combined data taken at
different times by virtue of the internal and stable noise source
(CM), whose signal was periodically (two times every 48$^m$) injected
towards the antenna port of the circulator in order to check the
SWR (\emph{reflectometer mode}) and, after that, towards the
receiver port as a calibration mark\footnote{The calibration mark
is the value above the noise floor of the signal injected by the
noise generator, expressed in analog-to-digital units (ADU). The
gain is properly the conversion factor K/ADU.}. In this way, we
could associate a value of the CM to every set of observations.
The different sets were then normalized to the reference CM
measured during absolute calibrations. We have been able to find
time sequences of good data (good weather conditions, no radio
interferences) lasting more than 48 hours. This allowed to compare
nighttime data separated up to two sidereal days and correct for
linear drifts of the signal (if present). The nighttime data,
corrected for drift and renormalized by means of the CM, have been
seamed at common RA. Integration of all the nighttime profiles
acquired during the life of the experiment gives the final
profile. Its scale, in ADU and arbitrary zero level, was then
converted to temperature scale and zero adjusted fitting the
profile to the absolute values of sky temperature measured during
calibration sessions. The reconstruction of the full sky profiles
at 0.6 GHz and 0.82 GHz deserves some other comments. The
calibration mark signals have a very small dispersion, witnessing
the stability of the system all along the observational campaign.
In fact the sample collected during several months at 0.6 GHz had
a mean value around 24000 ADU, while the mean standard deviation
set around 20 ADU. The corresponding quantities at 0.82 GHz were
respectively 19500 ADU and 30 ADU.

\noindent The error budget on the two definitive profiles at 0.6
and 0.82 GHz are then dominated by the uncertainties due to the
correction for the linear drifts and offsets and by the
uncertainty in the determination of the temperature scale. In
particular, along the galactic halo region ($8^h \lesssim \alpha
\lesssim 16^h$), which has been observed redundantly, the
definitive uncertainty upon temperature variations are in the
range $5 - 10$ mK both at 0.6 GHz and 0.82 GHz, depending on the
sky position (this uncertainty rises to $30 - 40$ mK near the
galactic plane, where we have collected a smaller number of drift
scans).

By this procedure we recover the full dynamic of the celestial
signal up to the antenna mouth and above the earth atmosphere.
Results are shown in Figure \ref{F5:TRISfinal}. This procedure
worked successfully at 0.6 and 0.82 GHz. At 2.5 GHz the
construction of drift scan profile was hampered by the
interference level which frequently covered the sky signal. The
few observing hours when interferences were small are so sparse
that we couldn't build a clean drift scan profile.

\section{Discussion}\label{Discussion}
The absolute values of $T_{sky}(\alpha,\delta)$ measured by TRIS
at 0.6005 GHz, 0.8178 GHz and 2.4278 GHz, at selected points of
constant declination $+42^{\circ}$ are listed in Tables
\ref{tab7}, \ref{tab8} and \ref{tab9}. The same data are plotted
in Figure \ref{F5:TRISfinal} together with the drift scans (0.6
GHz and 0.82 GHz are reported in Tables \ref{tab10} and
\ref{tab11}). Having no reliable drift scan at 2.5 GHz, we built a
profile only of the galactic component starting from the Stockert
\cite[]{Reich1986} survey at 1.420 GHz. This map has been
convoluted to the beam of the TRIS antennas in order to get the
synthetic drift scan at $\delta =+42^\circ$. Then we compared this
scan with the pure galactic signal we evaluated at 0.6 and 0.82
GHz (see Papers II and III). Using these measurements we evaluated
the galactic emission first at 1.420 GHz and then extrapolated it
at 2.5 GHz, using the local spectral index evaluated between 0.6
and 0.82 GHz. We used the local spectral index because it is a
well defined quantity throughout all the sky provided that
absolute galactic emissions measurements are available, which is
our case. We used the map at 1.42 GHz because it is the closest
one, public domain, at the frequency 2.5 GHz. In Figure
\ref{F5:TRISfinal} at 2.5 GHz, for uniformity with the measured
sky profiles plotted at 0.6 and 0.82 GHz, we added over the TRIS
absolute sky temperatures a profile (dashed line) obtained summing
to the galactic component calculated as described above, the
Unresolved Extragalactic Radio Sources contribution
\cite{Gervasi2008b} and Cosmic Microwave Background
\cite{Gervasi2008a}.

The systematic uncertainties are computed propagating through
Equation (\ref{eq:Tant}) and (\ref{eq:Tsky}) the uncertainties on
the measured values of attenuation and temperature of the lossy
components between the switch and the antenna mouth. The losses are very small. In some cases they are so small that
measurements give only upper limits (see Table \ref{tab4}) and the range of
fluctuation of the true losses is unknown but reasonably smaller and well
inside the upper limit, set by the accuracy of the instrumentation used to
carry on the measurements. The final uncertainties on the sky temperature obtained
propagating these errors are therefore considered as systematic.
At 0.6 GHz the great majority of the components is cold ($\sim 4$
K) and the final systematic uncertainty is small (66 mK). At 2.5
GHz the systematic uncertainty is worse because of the quality of
the e.m. matching of the SPTT switch and the presence of an
additional section of warm waveguide between the dewar flange and
the antenna throat. This correction contributes to the systematics
for 284 mK. At 0.82 GHz the systematic uncertainty is very large,
660 mK, essentially because all the components above the switch
were warm and at high temperature (see Equation
\ref{eq:DeltaTnoise}). Such a large value, however, does not take
into account constraints set by unphysical situations (e.g.
negative values of the temperature of the diffuse radiation
components and values of the galactic spectral index not supported
by models of galactic emission). When astrophysical assumptions
are made on these values (for a complete discussion see Paper III)
the systematic uncertainty at 0.82 GHz is reduced to
$(^{+430}_{-300})$ mK.

The above results hold if the sky radiation is unpolarized. We
know however that the galactic component of $T_{sky}$ is partially
polarized and that polarization is linear. Since antennas are
polarized, our measured values of the sky temperature are better
written as
\begin{equation}
T_{sky}(\alpha,\delta,\nu) = T_{CMB}(\nu) + T_{uers}(\nu) +
T_{gal}^{unpol}(\alpha,\delta,\nu) +
T_{gal}^{pol}(\alpha,\delta,\nu)~\cos^2 [\theta(\alpha,\delta,\nu)
- \theta_0] \label{eq:Tpol}
\end{equation}

\par\noindent where $T_{gal}^{unpol}$ and $T_{gal}^{pol}$ are
respectively the unpolarized and polarized components of the
galactic signal, and ($\theta - \theta_0$) is the angle between
the polarization vector of the radiation and the polarization
vector of the antenna.
\par\noindent The true sky temperature is therefore
\begin{equation}
T_{sky}^{true}(\alpha,\delta,\nu) =
T_{sky}^{unpol}(\alpha,\delta,\nu) +
T_{sky}^{pol}(\alpha,\delta,\nu) = T_{sky}(\alpha,\delta,\nu) +
\Delta T^{pol}(\alpha,\delta,\nu) \label{eq:Ttrue}
\end{equation}
\par\noindent where
\begin{equation}
\Delta T^{pol}(\alpha,\delta,\nu) =
T_{gal}^{pol}(\alpha,\delta,\nu)~\sin^2 [\theta(\alpha,\delta,\nu) -
\theta_0] \label{eq:DeltaT}
\end{equation}
\par\noindent is the signal lost. Were circular
polarization present, an additional term should be added to
Equation (\ref{eq:Ttrue}).

The information today available on the polarization degree and
direction of the diffuse radiation is limited. We do not expect
circular polarization. Linear polarization on the contrary must be
expected but the information available is poor. Below 1 GHz there
are only data obtained thirty years ago by \cite{Brouw1976}. The
assessment of the overall sky temperature has been obtained
correcting our absolute measurements for the fraction of the
linearly polarized signal. In order to do this we used the Brouw
and Spoelstra maps of the Stokes Parameters Q$\&$U in the
resampled version prepared by E.Carretti in the framework of the
SPOrt program \cite{Cortiglioni 2004}. Carretti maps having
7$^\circ$ resolution, were further convolved with the beam of our
experiment. Then, considering the relative orientation of the
polarization plane (E-plane) of our antennas and the polarization
of the celestial signal, we projected the polarized vector into
Copolar (collected by our antennas) and Crosspolar components
(rejected by our antennas). We conclude that, at TRIS angular resolution, the maximum
contribution of the polarized signal to the overall one, around
2$\%$ at 0.6 GHz and $3\%$ at 0.82 GHz, comes mostly from regions
far from the galactic disk. This result is in agreement with
Platania et al. (1998) who gives a maximum contribution of $\sim
5\%$ at 0.408 GHz with a $18^\circ$ beam in a sky region very close
to the one scanned by TRIS. A detailed discussion can be found in
Paper III. Between 1 and 3 GHz more recent and accurate data are
available (see for instance
 \cite{Duncan1991}, \cite{Wolleban2006} and references therein), but with
few exception they do not fill all the areas covered by TRIS
scans. Full sky models of the galactic synchrotron intensity and
linear polarization prepared for feasibility studies of
forthcoming space experiments (e.g. \cite{Giardino2002},
\cite{Bernardi2003}) are available. However we preferred to
evaluate the polarized component starting from real data acquired
at frequencies close to the TRIS ones. Considering that at 2.5 GHz
with the TRIS angular resolution
$(T_{gal}/T_{sky})\simeq(T_{gal}/T_{CMB})\leq 0.1$ and $T_{gal}^{pol}
\leq T_{gal}$, we can assume also at 2.5 GHz $(\Delta
T^{pol}/T_{sky}^{true})<0.02$.

In conclusion, neglecting the polarization effects is equivalent
to introduce an additional systematic uncertainty on the true
values of the sky temperature of less than $3\%$ (see Table
\ref{tab12}).

Table \ref{tab13} gives a list of measurements of $T_{Sky}$
present in literature at frequencies comparable to the frequencies
used by TRIS, the value of the derived $T_{CMB}$ and the
associated uncertainties for $T_{Sky}$ and $T_{CMB}$. All the
values of $T_{Sky}$ have been obtained as intermediate steps of
experiments aimed at getting the CMB temperature, so papers
usually focus the reader attention on the accuracy $\sigma_{CMB}$
of $T_{CMB}$. The accuracy $\sigma_{sky}$ of $T_{sky}$ should be
better ($\sigma_{sky}~<~\sigma_{CMB}$) but extracting it from
papers it is not always possible.  When no detailed information is
available in literature we assume $\sigma_{sky} ~\leq~
\sigma_{CMB}$. It appears that at 0.6 and 0.82 GHz the TRIS
absolute values of the sky temperature are the best today
available. At 2.5 GHz TRIS results are less accurate than data
obtained in the past by us with the White Mt. and the South Pole
collaborations. Nevertheless we report also the TRIS results at
2.5 GHz because they came exactly from the same direction and have
been obtained with the same angular resolution of our observation
at lower frequencies.

\noindent TRIS data can be used to

\noindent i) disentangle the components of the diffuse radiation,

\noindent ii) extract the cosmological and astrophysical
information carried by these components,

\noindent iii) improve zero level and scale of temperature of the
full sky maps of the diffuse radiation at decimetric wavelength in
literature (see Tables \ref{tab1} and \ref{tab2}).

These types of analysis are intricate and require different
approaches, depending on the final aim: astrophysical or
cosmological conclusion. They are done in separate Papers II
\cite{Gervasi2008a}, III \cite{Tartari2008} and
\cite{Gervasi2008b} which accompany present Paper I. Here (Table
\ref{tab14}) we present a summary of the results discussed in the
cited papers. We have reduced the error bars on $T_{CMB}$ of a
factor 9 at $\nu=0.6$ GHz and of a factor 7 at $\nu=0.82$ GHz.
These results have been possible because of the improvements in
the absolute calibration system and in the foregrounds separation
technique. At 2.5 GHz TRIS did not improve the previous
measurements, but is in agreement with them.

\acknowledgments  {\bf Acknowledgements}: The TRIS activity has
been supported by MIUR (Italian Ministry of University and
Research), CNR (Italian National Council of Research) and the
Universities of Milano and of Milano-Bicocca. The logistic support
at Campo Imperatore was provided by INFN, the Italian Institute of
Nuclear Physics, and its Laboratorio Nazionale del Gran Sasso. We
are indebted with E. Carretti who provided us a low resolution
polarized map based on the Brouw and Spoelstra one. We are also
grateful to R. Nesti for some crucial simulations of the horn
attenuations. We acknowledge the contributions and comments by
many peoples, among them G. Bonelli, A. Grillo, C. Cattadori, G.
De Amici, E. Pagana, G. Navarra, of members of the LNGS technical
staff A. Fulgenzi, G. Adinolfi , L. Masci, G. Giuliani. Frequently
students who were preparing their thesis for the degree in Physics
took part to TRIS activities. We acknowledge the use of the
HEALPix package \cite{Gorski2005} to evaluate the polarized
galactic signal at 0.610 and 0.82 GHz.


\clearpage

\begin{figure}[b]
\begin{center}
\includegraphics[angle=0,scale=.75]{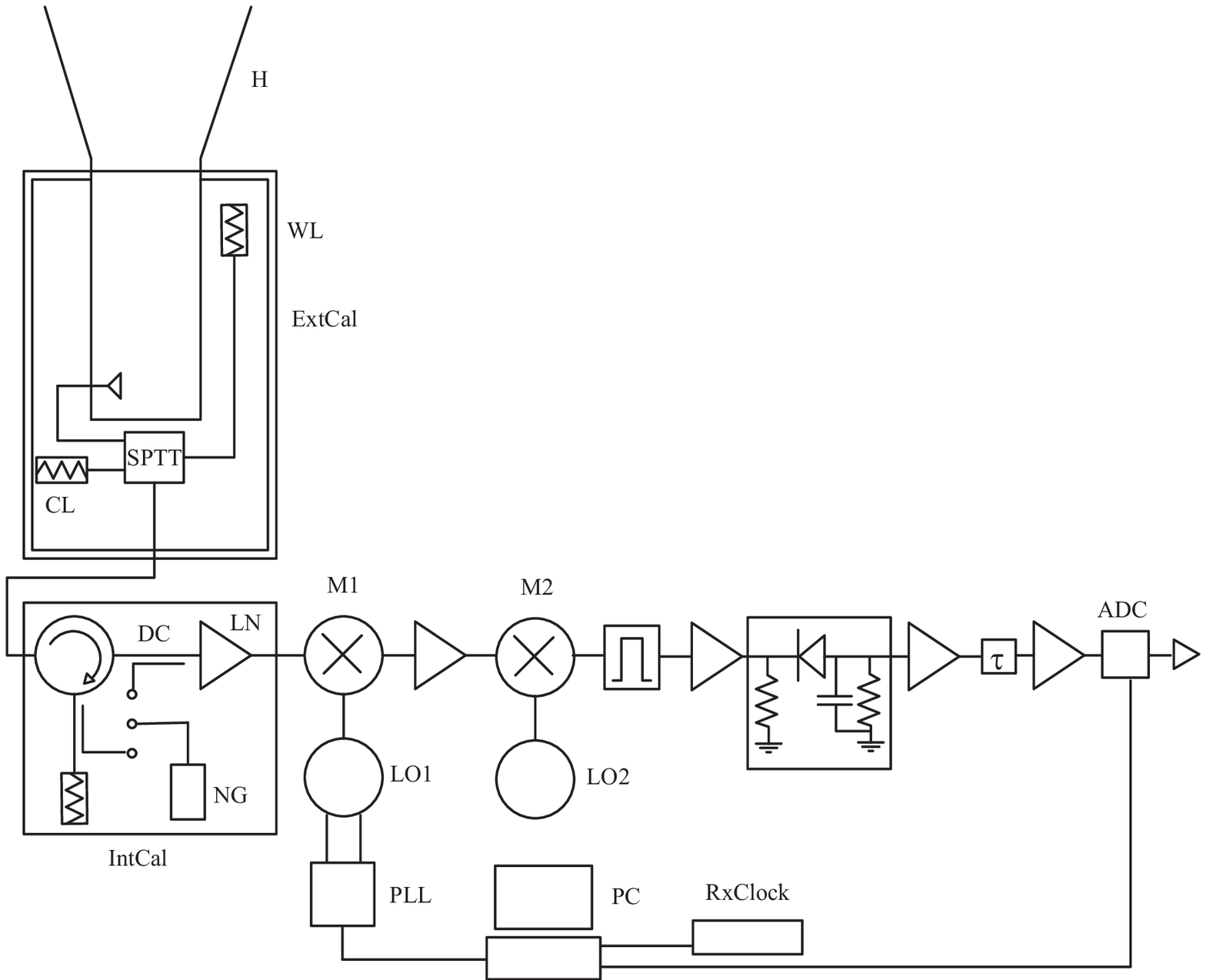}
\caption{Block diagram of TRIS Radiometers  ~H = corrugated Horn,
~LN = Low Noise Amplifier,  ~LO1, LO2 = Local Oscillators,  ~M1,
M2 = mixers, ~PLL = Phase Locked Loop, ~~${\bf \tau}$ = system
time constant, ~ADC = analog to digital converter, ~PC = Personal
Computer, ~RxClock = Radio Clock,  ~ExtCal = External Calibrator,
(WL = {\it warm Load}, ~CL = {\it Cold Load}, ~SPTT = {\it
switch}), ~IntCal = Internal calibrator ~( C = {\it Circulator},
~DC = {\it Directional Coupler}, ~NG ={\it Noise Generator})}
\label{F1:block}
\end{center}
\end{figure}

\begin{figure}[b]
\begin{center}
\includegraphics[angle=0,scale=.50]{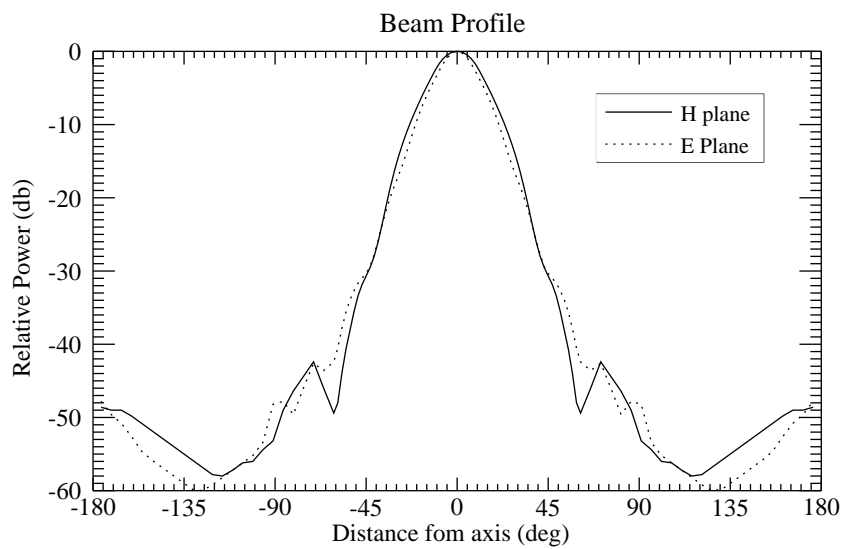}
\caption{Beam of TRIS antennas measured on a geometrically scaled
model at 8.2 GHz (continuous line: H-plane, dotted line:
E-plane)}\label{F2:beam}
\end{center}
\end{figure}

\begin{figure}[b]
\begin{center}
\includegraphics[angle=0,scale=.50]{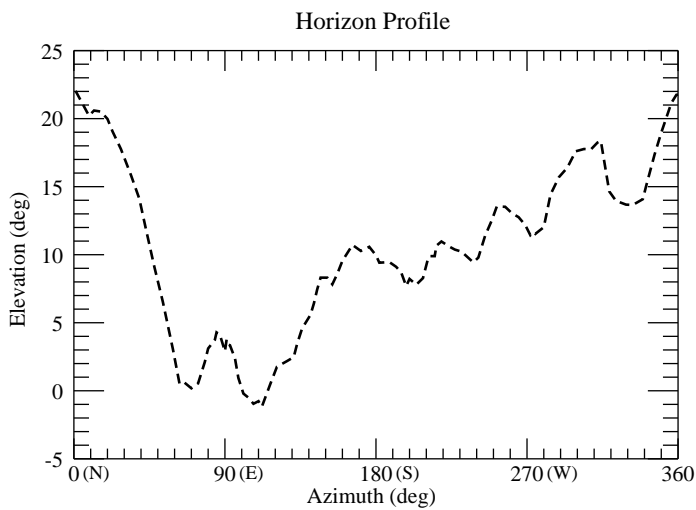}
\caption{Horizon profile at Campo Imperatore}\label{F3:horiz}
\end{center}
\end{figure}

\begin{figure}[b]
\begin{center}
\includegraphics[angle=0,scale=.50]{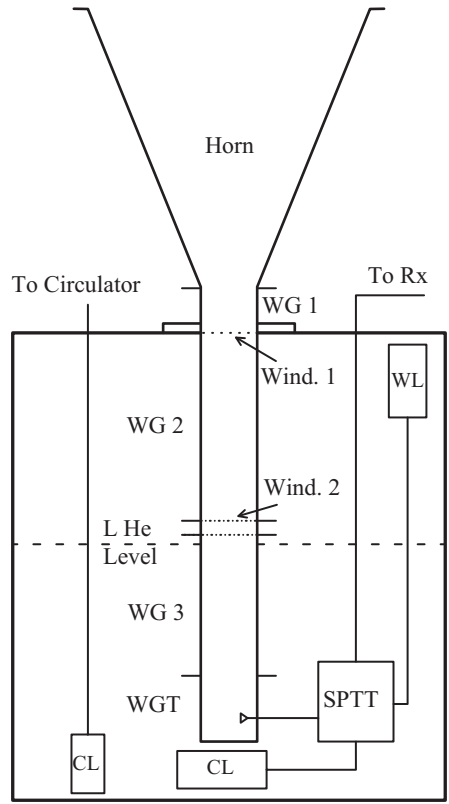}
\caption{A schematic view of the Cryogenic External Calibrator (WG
1 = anticorodal waveguide section, WG 2, WG 3 = stainless steel
waveguide sections, Wind. 1 = polyethylene window, Wind. 2 =
fluorglass windows, WGT = brass coax-waveguide transition, SPTT =
Single Pole Triple Throw switch, WL = warm dummy load, CL = cold
dummy load; see also Table \ref{tab4} and text)} \label{F4:dewar}
\end{center}
\end{figure}

\begin{figure}[b]
\begin{center}
\includegraphics[angle=0,scale=1.0]{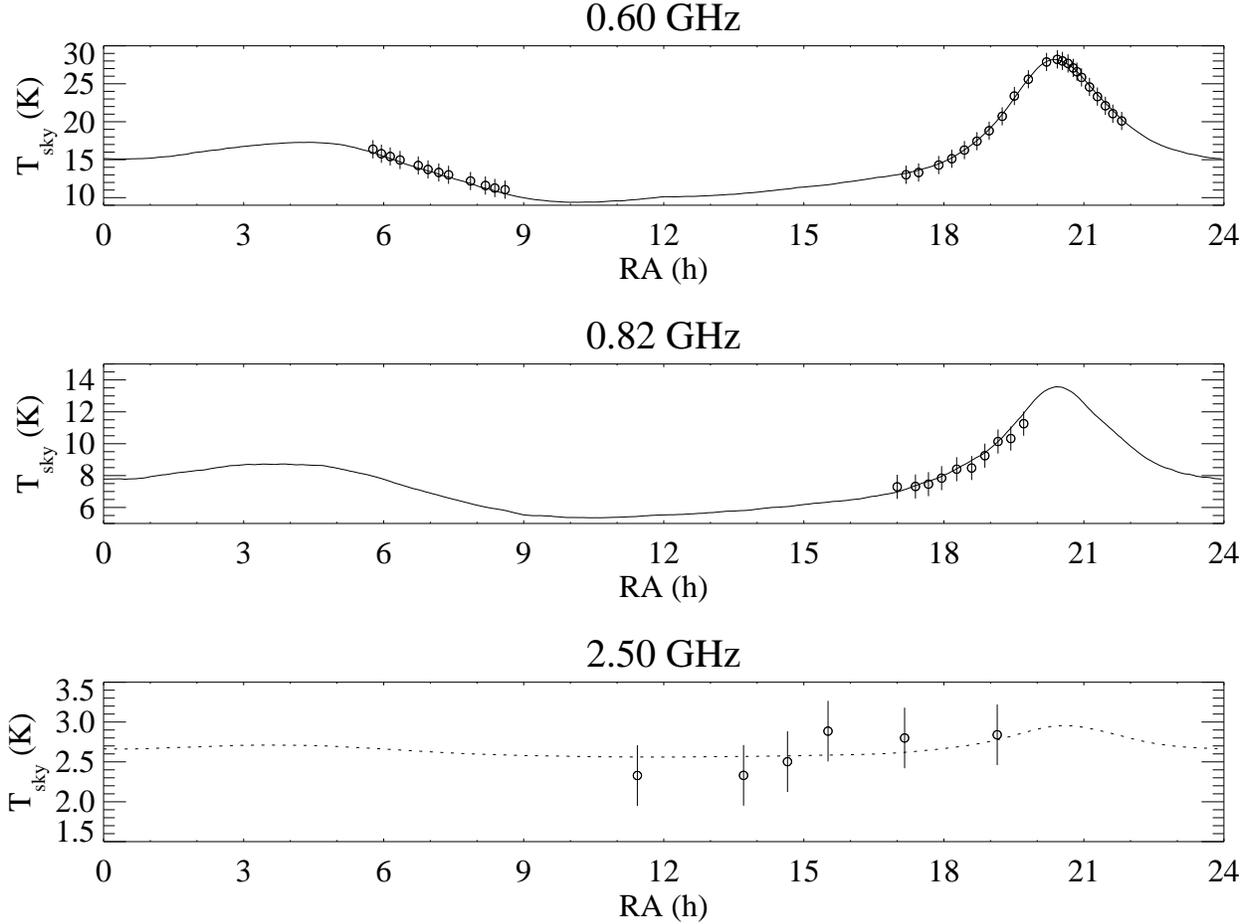}
\caption{Absolute values and differential profiles of $T_{sky}$
measured by TRIS  at $\delta = +42^{\circ}$, ~($0~\leq \alpha
\leq~24$). Uncertainties (3$\sigma$ statistic + systematic)
associated to the absolute values are indicated: to make them
visible at 0.60 GHz the error bars have been enlarged by a factor
10. The zero level of the differential profiles has been adjusted
to fit the distribution versus $\alpha$ of the absolute values. At
2.50 GHz the plotted sky profile (dashed line) doesn't come from
TRIS data but it has been obtained rescaling the sky profile
extracted from the Stockert \cite{Reich1986} survey at 1.4 GHz. See Paper III
for details.}\label{F5:TRISfinal}
\end{center}
\end{figure}

\clearpage
\begin{deluxetable}{lccccr}
\tablewidth{0pt} \tabletypesize{} \tablecolumns{6}
\tablecaption{Extended maps of the diffuse radiation}
\tablehead{\colhead{$\nu$ (GHz)} & \colhead{$\delta T_{zero}$ (K)}
& \colhead{$\delta T$} & \colhead{Angular Res.} & \colhead{Sky
Coverage} & \colhead{Reference}} \startdata
0.022&$\simeq 10^3$&n.a.&$1.1^{\circ} \times 1.7^{\circ}$& North& 1\\
0.038&300&$5 \%$&$7.5^{\circ}$&North&2 \\
0.151&40&$\simeq 10
\%$&$5^{\circ}$&full&3\\
0.408&3&$\simeq 10\%$&$0.85^{\circ}$&full&4\\
0.820&0.6&$\simeq 10 \%$&$1.2^{\circ}$&North&5\\
1.420&0.5&$\simeq 10 \%$&$0.6^{\circ}$&North&6\\
2.326&0.08&$\leq 5 \%$&$0.33^{\circ}$&South&7\\
\enddata
\tablerefs{(1) Roger et al. 1999; (2) Milogradov-Turin 1984; (3)
Landecker \& Wielebinsky 1970; (4) Haslam et al. 1982; (5)
Berkhuijsen 1972; (6) Reich \& Reich 1986; (7) Jonas et al. 1998}
\label{tab1}
\end{deluxetable}

\clearpage
\begin{deluxetable}{cc}
\tablewidth{0pt} \tabletypesize{} \tablecolumns{2}
\tablecaption{Typical accuracy of measured values of $T_{sky}$ in
literature} \tablehead{\colhead{$\nu_{0}$ (GHz)} &
\colhead{$\sigma(mK)$}} \startdata
100~--~60~& 1~--~10 \\
~60~--~10~& 10~--~50 \\
~10~--~~1~& 10~--~500 \\
~1~--~0.5~&~500~--~1500
\enddata
\label{tab2}
\end{deluxetable}

\clearpage
\begin{deluxetable}{lccc}
\tablewidth{0pt} \tabletypesize{} \tablecolumns{4}
\tablecaption{TRIS antennas} \tablehead{\colhead{$\nu_0$ (GHz)} &
\colhead{0.60} & \colhead{0.82} & \colhead{2.5}} \startdata Horn
Aperture & $3.7\lambda\times4.9\lambda$ &
$3.7\lambda\times4.9\lambda$ & $3.7\lambda\times4.9\lambda$ \\
Flare Angle E Plane & $19^{\circ}$ & $19^{\circ}$ & $19^{\circ}$ \\
Flare Angle H Plane & $23^{\circ}$ & $23^{\circ}$ & $23^{\circ}$ \\
Phase difference $\delta_E$ &$0.07\lambda$&$0.07\lambda$&$0.07\lambda$ \\
Phase difference $\delta_H$ &$0.10\lambda$&$0.10\lambda$&$0.10\lambda$ \\
HPBW  & $18^{\circ}(E)\times23^{\circ}(H)$ & $18^{\circ}(E)\times23^{\circ}(H)$ & $18^{\circ}(E)\times23^{\circ}(H)$\\
Mouth Dimensions (m)& $1.85\times2.41$ & $1.35\times1.79$ & $0.44\times0.59$ \\
Horn length (m) & 2.50 & 1.83 & 0.60 \\
Back Lobes (dB) & $< -40$ & $< -40$ & $< -40$
\enddata
\label{tab3}
\end{deluxetable}

\clearpage
\begin{deluxetable}{lccr}
\tablewidth{0pt} \tabletypesize{} \tablecaption{Transmission
$e^{-\tau}$ of the TRIS front end components} \tablecolumns{4}
\tablehead{\colhead{component\tablenotemark{\dag}} &
\colhead{$\nu=$ 0.6 GHz} & \colhead{$\nu=$ 0.82 GHz} &
\colhead{$\nu=$ 2.5 GHz}} \startdata
WG Cable (cold) & $ 0.957 \pm 0.001 $ & $ 0.9698 \pm 0.0004 $ & $ 0.969 \pm 0.006 $ \\
WG Cable (warm) & $ $\nodata$ $ & $ 0.9600 \pm 0.0001 $ & $ $\nodata$ $ \\
WG Transition   & $ 0.988 \pm 0.002 $ & $ 0.986 \pm 0.004 $ & $ 0.984 \pm 0.004 $ \\
WG3-1 & $ 0.9981 \pm 0.0003 $ & $ 0.9966 \pm 0.0002 $ & $ 0.997 \pm 0.002 $ \\
wind-1 & $ >0.99999 $ & $ >0.99999 $ & $> 0.99998 $ \\
WG3-2 & $ 0.9983 \pm 0.0003 $ & $ 0.9965 \pm 0.0002 $ & $ 0.998 \pm 0.001 $ \\
wind-2 & $ >0.99998 $ & $ >0.99998 $ & $ >0.99998 $ \\
WG2 & $ 0.9989 \pm 0.0002 $ & $ 0.9997 \pm 0.0002 $ & $ 0.9995 \pm 0.0003 $ \\
WG1 & $ 0.9983 \pm 0.0001 $ & $ 0.9991 \pm 0.0001 $ & $ 0.9985 \pm 0.0001 $ \\
WG ex & $ $\nodata$ $ & $ $\nodata$ $ & $ 0.9978 \pm 0.0002 $ \\
Horn & $ 0.9920 \pm 0.0001 $ & $ 0.9872 \pm 0.0001 $ & $ 0.9972 \pm 0.0001 $ \\
\enddata
\tablenotetext{\dag}{See text and Figure \ref{F4:dewar}}
\label{tab4}
\end{deluxetable}

\clearpage
\begin{deluxetable}{lccc}
\tablewidth{0pt} \tabletypesize{} \tablecaption{TRIS receiver}
\tablecolumns{4} \tablehead{\colhead{$\nu_0$ (GHz)} &
\colhead{0.60} & \colhead{0.82} & \colhead{2.5}} \startdata
Tunability frequency range (MHz) & 20 & 20 & 200 \\
Number of selectable frequencies &256 & 256 & 256 \\
Integration bandwidth (MHz) & 0.3 & 0.3 & 3.0 \\
Sampling time (s) & 4 & 4 & 4 \\
Noise Temperature (K) & 120 & 120 & 100 \\
$T_{RX}^{eff}$(K)\tablenotemark{\dag} & 84.5 & 67.3 & 72.6 \\
$T^{eff}_{cold}$(K)\tablenotemark{\dag} & 4.1 & 4.3 & 7.1 \\
$T^{eff}_{warm}$(K)\tablenotemark{\dag} & 231.4 & 247.2 & 247.1 \\
Gain ($10^{-4}$K/div) & $8.498 \pm 0.004$ & $7.130 \pm 0.001$ &
$7.234 \pm 0.001$
\enddata
\tablenotetext{\dag}{Typical values (helium level dependent)}
\label{tab5}
\end{deluxetable}

\clearpage
\begin{deluxetable}{lrrc}
\tablewidth{0pt} \tablecaption{Atmospheric and environment
contributions} \tablecolumns{4} \tablehead{\colhead{$\nu_{0}$} &
\colhead{0.60 GHz} & \colhead{0.82 GHz} & \colhead{2.5 GHz}}
\startdata
$ T_{atm} $ (K) & $ 1.088 \pm ~0.023 $ & $ 1.221 \pm 0.016 $ & $ 1.570 \pm 0.025 $ \\
$ e^{-\tau_{atm}} $ & $ 0.995 \pm 0.001 $ & $ 0.995 \pm 0.001 $ & $ 0.993 \pm 0.001 $\\
$ T_{ground} $ (K) & $ 0.07^{+0.06}_{-0.03} $ & $ 0.07^{+0.06}_{-0.03} $ & $ 0.07^{+0.06}_{-0.03} $ \\
$ T_{interf} $ (K) & $ <0.01 $ & $ <0.01 $ & $ 9.82 \pm 0.26 $ \\
\enddata
\label{tab6}
\end{deluxetable}

\clearpage
\begin{deluxetable}{cccccccc}
\tablewidth{0pt} \tablecaption{Absolute values of $T_{sky}$ at
$\nu =$ 0.60 GHz, $\delta = +42^{\circ}$} \tablecolumns{8}
\tablehead{\colhead{$~\alpha (h,m,s)$} & \colhead{$T_{sky}$
(K)\tablenotemark{\dag}} & \colhead{~~~} & \colhead{$~\alpha
(h,m,s)$} & \colhead{$T_{sky}$ (K)\tablenotemark{\dag}}  &
\colhead{~~~} & \colhead{$~\alpha (h,m,s)$} & \colhead{$T_{sky}$
(K)\tablenotemark{\dag}}} \startdata
05 46 00    &   16.372  &&  17 11 18    &   13.016  &&  20 32 35    &   28.007  \\
05 57 02    &   15.788  &&  17 27 31    &   13.315  &&  20 39 18    &   27.678  \\
06 08 26    &   15.423  &&  17 53 37    &   14.292  &&  20 45 59    &   27.084  \\
06 20 46    &   14.962  &&  18 10 12    &   15.123  &&  20 50 46    &   26.592  \\
06 44 22    &   14.241  &&  18 26 23    &   16.267  &&  20 57 11    &   25.839  \\
06 57 14    &   13.717  &&  18 42 33    &   17.437  &&  21 07 13    &   24.582  \\
07 10 34    &   13.319  &&  18 58 24    &   18.813  &&  21 17 17    &   23.312  \\
07 23 18    &   13.010  &&  19 14 56    &   20.715  &&  21 27 18    &   22.100  \\
07 51 39    &   12.193  &&  19 30 39    &   23.386  &&  21 37 20    &   21.061  \\
08 10 34    &   11.602  &&  19 48 22    &   25.593  &&  21 48 40    &   20.107  \\
08 22 56    &   11.264  &&  20 12 02    &   27.869  &&      &       \\
08 36 02    &   11.050  &&  20 25 52    &   28.211  &&      &       \\
\enddata
\tablenotetext{\dag}{For error bars see Table \ref{tab12}.}
\label{tab7}
\end{deluxetable}

\clearpage
\begin{deluxetable}{cccccccc}
\tablewidth{0pt} \tablecaption{Absolute values of $T_{sky}$ at
$\nu = 0.82$ GHz, $\delta = +42^{\circ}$} \tablecolumns{8}
\tablehead{\colhead{$~\alpha (h,m,s)$} & \colhead{$T_{sky}$
(K)\tablenotemark{\dag}} & \colhead{~~~} & \colhead{$~\alpha
(h,m,s)$} & \colhead{$T_{sky}$ (K)\tablenotemark{\dag}}  &
\colhead{~~~} & \colhead{$~\alpha (h,m,s)$} & \colhead{$T_{sky}$
(K)\tablenotemark{\dag}}} \startdata
17 00 04    &   7.297   &&  18 16 29    &   8.401   &&  19 26 04    &   10.322  \\
17 23 29    &   7.319   &&  18 35 40    &   8.476   &&  19 42 39    &   11.256  \\
17 40 15    &   7.463   &&  18 52 43    &   9.249   &&  \\
17 57 06    &   7.840   &&  19 09 38    &   10.137  &&  \\
\enddata
\tablenotetext{\dag}{For error bars see Table \ref{tab12}}
\label{tab8}
\end{deluxetable}

\clearpage
\begin{deluxetable}{cccccccc}
\tablewidth{0pt} \tablecaption{Absolute values of $T_{sky}$ at
$\nu = 2.5$ GHz, $\delta = +42^{\circ}$} \tablecolumns{8}
\tablehead{\colhead{$~\alpha (h,m,s)$} & \colhead{$T_{sky}$
(K)\tablenotemark{\dag}} & \colhead{~~~} & \colhead{$~\alpha
(h,m,s)$} & \colhead{$T_{sky}$ (K)\tablenotemark{\dag}}  &
\colhead{~~~} & \colhead{$~\alpha (h,m,s)$} & \colhead{$T_{sky}$
(K)\tablenotemark{\dag}}} \startdata
11 26 04    &   2.329  & & 14 39 01    &   2.503  & & 17 09 34    &   2.800  \\
13 42 32    &   2.331  & & 15 31 06    &   2.886  & & 19 08 35    &   2.840  \\
\enddata
\tablenotetext{\dag}{For error bars see Table \ref{tab12}}
\label{tab9}
\end{deluxetable}

\clearpage
\begin{deluxetable}{cccccccc}
\tablecaption{Sampled values of $T_{sky}$ profile at $\nu =$ 0.60
GHz, $\delta = +42^{\circ}$} \tablecolumns{8}
\tablehead{\colhead{$~\alpha (h,m,s)$} & \colhead{$T_{sky}$
(K)\tablenotemark{\dag}} & \colhead{~~~} & \colhead{$~\alpha
(h,m,s)$} & \colhead{$T_{sky}$ (K)\tablenotemark{\dag}}  &
\colhead{~~~} & \colhead{$~\alpha (h,m,s)$} & \colhead{$T_{sky}$
(K)\tablenotemark{\dag}}} \tablewidth{0pt} \startdata
00 00 00    &   15.145  &&  08 00 00    &   11.552  &&  16 00 00    &   12.117  \\
00 30 00    &   15.099  &&  08 30 00    &   10.683  &&  16 30 00    &   12.585  \\
01 00 00    &   15.204  &&  09 00 00    &   9.996   &&  17 00 00    &   13.023  \\
01 30 00    &   15.494  &&  09 30 00    &   9.576   &&  17 30 00    &   13.703  \\
02 00 00    &   15.986  &&  10 00 00    &   9.390   &&  18 00 00    &   14.747  \\
02 30 00    &   16.343  &&  10 30 00    &   9.412   &&  18 30 00    &   16.543  \\
03 00 00    &   16.745  &&  11 00 00    &   9.572   &&  19 00 00    &   19.201  \\
03 30 00    &   17.046  &&  11 30 00    &   9.801   &&  19 30 00    &   23.044  \\
04 00 00    &   17.242  &&  12 00 00    &   10.124  &&  20 00 00    &   27.015  \\
04 30 00    &   17.275  &&  12 30 00    &   10.168  &&  20 30 00    &   28.048  \\
05 00 00    &   17.050  &&  13 00 00    &   10.273  &&  21 00 00    &   25.628  \\
05 30 00    &   16.395  &&  13 30 00    &   10.442  &&  21 30 00    &   22.210  \\
06 00 00    &   15.443  &&  14 00 00    &   10.719  &&  22 00 00    &   19.323  \\
06 30 00    &   14.368  &&  14 30 00    &   11.058  &&  22 30 00    &   17.342  \\
07 00 00    &   13.438  &&  15 00 00    &   11.397  &&  23 00 00    &   16.204  \\
07 30 00    &   12.506  &&  15 30 00    &   11.682  &&  23 30 00    &   15.474  \\
\enddata
\tablenotetext{\dag}{For error bars see Table \ref{tab12}}
\label{tab10}
\end{deluxetable}

\clearpage
\begin{deluxetable}{cccccccc}
\tablewidth{0pt} \tablecaption{Sampled values of $T_{sky}$ profile
at $\nu =$ 0.82 GHz, $\delta = +42^{\circ}$} \tablecolumns{8}
\tablehead{\colhead{$~\alpha (h,m,s)$} & \colhead{$T_{sky}$
(K)\tablenotemark{\dag}} & \colhead{~~~} & \colhead{$~\alpha
(h,m,s)$} & \colhead{$T_{sky}$ (K)\tablenotemark{\dag}}  &
\colhead{~~~} & \colhead{$~\alpha (h,m,s)$} & \colhead{$T_{sky}$
(K)\tablenotemark{\dag}}} \startdata
00 00 00    &   7.781   &&  08 00 00    &   6.145   &&  16 00 00    &   6.481   \\
00 30 00    &   7.781   &&  08 30 00    &   5.869   &&  16 30 00    &   6.707   \\
01 00 00    &   7.935   &&  09 00 00    &   5.537   &&  17 00 00    &   6.969   \\
01 30 00    &   8.143   &&  09 30 00    &   5.471   &&  17 30 00    &   7.486   \\
02 00 00    &   8.322   &&  10 00 00    &   5.371   &&  18 00 00    &   8.026   \\
02 30 00    &   8.529   &&  10 30 00    &   5.357   &&  18 30 00    &   8.783   \\
03 00 00    &   8.678   &&  11 00 00    &   5.387   &&  19 00 00    &   9.724   \\
03 30 00    &   8.718   &&  11 30 00    &   5.456   &&  19 30 00    &   11.198  \\
04 00 00    &   8.702   &&  12 00 00    &   5.536   &&  20 00 00    &   12.905  \\
04 30 00    &   8.648   &&  12 30 00    &   5.589   &&  20 30 00    &   13.555  \\
05 00 00    &   8.465   &&  13 00 00    &   5.679   &&  21 00 00    &   12.547  \\
05 30 00    &   8.141   &&  13 30 00    &   5.778   &&  21 30 00    &   11.141  \\
06 00 00    &   7.772   &&  14 00 00    &   5.897   &&  22 00 00    &   9.839   \\
06 30 00    &   7.307   &&  14 30 00    &   6.026   &&  22 30 00    &   8.823   \\
07 00 00    &   6.902   &&  15 00 00    &   6.185   &&  23 00 00    &   8.235   \\
07 30 00    &   6.512   &&  15 30 00    &   6.331   &&  23 30 00    &   7.924   \\
\enddata
\tablenotetext{\dag}{For error bars see Table \ref{tab12}}
\label{tab11}
\end{deluxetable}

\clearpage
\begin{deluxetable}{lcccr}
\tablewidth{0pt} \tablecaption{Accuracy of the absolute values of
$T_{sky}$ measured by TRIS at $\delta = + 42^\circ$}
\tablecolumns{5} \tablehead{\colhead{$\nu $ (GHz)} &
\colhead{0.60} & \colhead{0.82} & \colhead{2.5}} \startdata
\em{statistical uncertainty} & & & \\
stand. dev. $\sigma~$ (mK) & 104 & 160 & 25 \\
mean stand. dev. $\sigma_m$~(mK) & 18 & 32 & 10 \\
& & & \\
\em{systematic uncertainty} & & & \\
zero level (mK) & 66 & $^{+430}_{-300}$\tablenotemark{\dag}& 284 \\
polarization effect & $<2\%$& $<3\%$& $<2\%$ \\
\enddata
\tablenotetext{\dag}{See section \ref{Discussion}.} \label{tab12}
\end{deluxetable}

\clearpage
\begin{deluxetable}{lrrrrr}
\tablewidth{0pt} \tablecaption{Accuracy of measurements in
literature of $T_{CMB}$ and $T_{sky}$ at frequencies close to 0.6,
0.82 and 2.5 GHz} \tablecolumns{6} \tablehead{\colhead{$\nu$
(GHz)} & \colhead{$\lambda$ (cm)} & \colhead{$\sigma_{sky}$ (K)} &
\colhead{$T_{CMB}$ (K)}& \colhead{$\sigma_{CMB}$ (K)}&
\colhead{Reference}} \startdata
0.610 & 49.1 & 0.7 & 3.7 & 1.2 & 1 \\
0.6 & 50.0 & 0.9 & 3.0 & 1.2 & 2 \\
0.635 & 47.2 & $\sim$0.5 & 3.0 & 0.5 & 3 \\
0.82 & 36.6 & 1.5 & 2.7 & 1.6 & 4 \\
2 & 15 & $\leq$ 0.3 & 2.5 & 0.3 & 5 \\
2 & 15 & 0.10 & 2.55 & 0.14 & 6 \\
2.3 & 13.1 & ~~0.2~-~0.4 & 2.66 & 0.7 & 7 \\
2.5 & 12 & 0.15 & 2.5 & 0.21 & 4 \\
\enddata
\tablerefs{(1) Howell $\&$ Shakeshaft 1967; (2) Sironi et al.
1990; (3) Stankevich et al. 1970; (4) Sironi et al. 1991; (5)
Pelyushenko $\&$ Stankevich 1969; (6) Bersanelli et al. 1994; (7)
Otoshi \& Stelzried 1975} \label{tab13}
\end{deluxetable}

\clearpage
\begin{deluxetable}{lcc}
\tablewidth{0pt} \tablecaption{TRIS results summary}
\tablecolumns{3} \tablehead{\colhead{0.6 GHz} &
\colhead{$\alpha^{1}=10^{h}00^{m}$} &
\colhead{$\alpha^{2}=20^{h}24^{m}$}} \startdata

T$_{sky}$ (K) & $ 9.390 \pm 0.066~(syst) \pm 0.018~(stat)$ & $ 28.190 \pm 0.066~(syst) \pm 0.018~(stat)$ \\
T$_{gal}$ (K) & $ 5.72 \pm 0.07$ & $ 24.44 \pm 0.07$ \\
T$_{UERS}$ (K) & \multicolumn{2}{c}{$ 0.934 \pm 0.024$}\\
T$_{CMB}$ (K) & \multicolumn{2}{c}{$ 2.823 \pm 0.066~(syst) \pm
0.129~(MC\tablenotemark{\dag}) $} \\
T$_{CMB}^{th}$ (K) & \multicolumn{2}{c}{$ 2.837 \pm 0.066~(syst) \pm
0.129~(MC\tablenotemark{\dag}) $} \\
\tableline \\[1.5pt]0.82 GHz & $\alpha^{1}=10^{h}00^{m}$ &
$\alpha^{2}=20^{h}24^{m}$ \\
\tableline
T$_{sky}$ (K) & $ 5.37^{+0.46}_{-0.30}~(syst) \pm 0.03~(stat)$ & $ 13.57^{+0.46}_{-0.30}~(syst) \pm 0.03~(stat)$ \\
T$_{gal}$ (K) & $ 2.21 \pm 0.03$ & $ 10.38 \pm 0.03$ \\
T$_{UERS}$ (K) & \multicolumn{2}{c}{$ 0.408 \pm 0.010$}\\
T$_{CMB}$ (K) & \multicolumn{2}{c}{$ 2.783^{+0.430}_{-0.300}~(syst) \pm 0.051~(MC\tablenotemark{\dag})$} \\
T$_{CMB}^{th}$ (K) & \multicolumn{2}{c}{$ 2.803^{+0.430}_{-0.300}~(syst) \pm 0.051~(MC\tablenotemark{\dag})$} \\
\tableline \\[1.5pt]2.5 GHz & $\alpha^{1}=10^{h}00^{m}$ &
$\alpha^{2}=20^{h}24^{m}$ \\
\tableline
T$_{sky}$ (K) & $ 2.57 \pm 0.28~(syst) \pm 0.10~(stat)$ & $ 2.99 \pm 0.28~(syst) \pm 0.10~(stat)$ \\
T$_{gal}\tablenotemark{\dag\dag}$ (K) & $ 0.091 \pm 0.093~(syst) \pm 0.005~(stat)$ & $ 0.471 \pm 0.093~(syst) \pm 0.027~(stat)$ \\
T$_{UERS}$ (K) & \multicolumn{2}{c}{$ 0.022 \pm 0.001$}\\
T$_{CMB}$ (K) & \multicolumn{2}{c}{$ 2.458 \pm 0.284~(syst) \pm 0.139~(stat)$} \\
T$_{CMB}^{th}$ (K) & \multicolumn{2}{c}{$ 2.516 \pm 0.284~(syst) \pm 0.139~(stat)$} \\
\enddata
\tablenotetext{\dag}{This uncertainty was evaluated by means of
MonteCarlo simulations, as described in Paper II.}

\tablenotetext{\dag\dag}{Due to the incompleteness of the TRIS
drift scan at 2.5 GHz, here T$_{gal}$ is extrapolated by (Reich \&
Reich 1986) map at 1.42 GHz convolved with TRIS beam and using
local spectral index calculated from our data at 0.6 and 0.82 GHz.
The quoted systematic uncertainty for T$_{gal}$ is relative to the
determination of the galactic signal at 1.42 GHz, starting from
the absolute measurements at 0.6 and 0.82 GHz.} \label{tab14}
\end{deluxetable}


\begin{thebibliography}{}
\bibitem[Ajello et al. 1995]{Ajello1995} Ajello, C., Bonelli, G.,
Sironi, G., 1995, \apjs ~96, 643
\bibitem[Bensadoun et al. 1992]{Bensadoun1992} Bensadoun, M., Witebsky, C., Smoot, G., de Amici, G.,
Kogut, A., Levin, S., 1992, RScI 63, 4377-4389
\bibitem[Bensadoun et al. 1993]{Bensadoun1993} Bensadoun, M., Bersanelli, M.,
de Amici, G., Kogut, A., Levin, S.~M., Limon, M., Smoot, G.~F.
Witebsky, C., 1993, \apj, 409, 1
\bibitem[Berkhuijsen 1972]{Berkhuijsen1982} Berkhuijsen, E.M., 1972, \aap, ~5, 263
\bibitem[Bernardi et al. 2003]{Bernardi2003} Bernardi, G., Carretti,
E., Cortiglioni, S., Sault, R.J., Kesteven, M.J., Poppi, S. 2003,
\apj, ~594, L5
\bibitem[Bersanelli et al. 1993]{Bersanelli1993} Bersanelli, M., Bonelli,G., Sironi, G. et al., 1993,
Antarctic Journal XXVIII, 306
\bibitem[Bersanelli et al. 1994]{Bersanelli1994} Bersanelli, M., Bensadoun, M., de Amici, G.,
Levin, S., Limon, M., Smoot, G. F., Vinje, W., 1994, \apj 424, 517
\bibitem[Brouw \& Spoelstra 1976]{Brouw1976} Brouw, W.N. and
Spoelstra, T.A.T., 1976, A\&A Suppl., ~26, 129
\bibitem[Burigana et al. 1991]{Burigana1991} Burigana, C. and Danese, L. and de Zotti, G., 1991, \aap, 246,
49B
\bibitem[Cortiglioni et al. 2004]{Cortiglioni 2004} Cortiglioni, S. et al., 2004, \na, 9, 297
\bibitem[Duncan et al. 1999]{Duncan1991} Duncan, A.R., Reich, P.,
Reich, W., Fuerst, E., 1999, \aap, ~350, 447
\bibitem[Fixsen \& Mather 2002]{Fixsen2002} Fixsen, D.J \& Mather, J.C. 2002,
\apj, ~581, 817
\bibitem[Gervasi et al. 2008a]{Gervasi2008a} Gervasi, M., Zannoni, M., Tartari, A.,
Boella, G., Sironi, G., 2008a, \apj, {\em submitted}
\bibitem[Gervasi et al. 2008b]{Gervasi2008b} Gervasi, M., Tartari, A., Zannoni, M.,
Boella, G., Sironi, G., 2008b, \apj, {\em in press},
arXiv:0803.4138v1
\bibitem[Giardino et al. 2002] {Giardino2002} Giardino, G., Banday, A.J., Gorsky,
K.M., Bennet, K., Jonas, J.L., Tauber, J.,  2002, \aap ~387, 82
\bibitem[Gorski et al. 2005]{Gorski2005} Gorski, K.~M., Hivon, E., Banday,
A.~J., Wandelt, B.~D., Hansen, F.~K., Reinecke, M., Bartelmann,
M., 2005, \apj, 622, 759
\bibitem[Haslam et al. 1982]{Haslam1982} Haslam, C.~G.~T., Salter, C.~J., Stoffel, H. and
Wilson, W.~E., 1982, \aaps, ~47, 1
\bibitem[Howell \& Shakeshaft 1967]{Howell1967} Howell, T.F. \& Shakeshaft,
J.R., 1967, Nature, 216, 7
\bibitem[Jonas et al. 1998]{Jonas1998} Jonas, J., Baart, E.E. and
Nicolson, G.D., 1998, MNRAS, ~297, 977
\bibitem[Kogut et al. 2004]{Kogut2004} Kogut,.a., Fixsen, D.J.,
Levin, S., Limon, M., Lubin, P.M., Mirel, P., Seiffert, M. and
Wollack, E. 2004, \apj Suppl. ~154, 493
\bibitem[Kogut et al. 2006]{Kogut2006} Kogut, A., Fixsen, D., Fixsen, S., Levin, S., Limon, M., Lowe,
L., Mirel, P., Seiffert, M., Singal, J., Lubin, P., Wollack, E.,
2006, \nar, 50, 925
\bibitem[Landecker \& Wielebinsky 1970]{Landecker1970} Landecker,
T.L. \& Wielebinsky, R., 1970, Austr. J. Phys. Suppl., ~16, 1
\bibitem[Mather et al. 1994]{Mather1994} Mather, J. C. et al,
1994, \apj ~420, 439
\bibitem[Milogradov-Turin 1984]{Milogradov1984} Milogradov-Turin, J.,
1984, \mnras, ~161, 269
\bibitem[Otoshi \& Stelzried 1975]{Otoshi1975} Otoshi, T.Y. \& Stelzried,
C.T., 1975, IEEE Trans. Instrum. Meas. 24, 174
\bibitem[Pelyushenko \& Stankevich 1969]{Pelyushenko1969} Pelyushenko, S.A. and Stankevich, K.S., 1969, Sov. Ast. 13, 223
\bibitem[Platania et al.(1998)]{Platania1998} Platania, P., Bensadoun, M., Bersanelli, M., De Amici, G., Kogut, A., Levin, S., Maino, D. \& Smoot, G.F.,
1998, \apj, 505, 473
\bibitem[Platania et al. 2003]{Platania2003} Platania, P., Burigana, C.,
Maino, D., Caserini, E., Bersanelli, M., Cappellini, B. and
Mennella A., 2003, \aap, ~410, 847
\bibitem[Reich \& Reich 1986]{Reich1986} Reich, P. \& Reich, W., 1986, \aap S, 63, 205
\bibitem[Roger et al. 1999]{Roger1999} Roger, R.S., Costain, C.H., Landecker, T.L. \& Swerdlyk, C.M., 1999, \aaps, 137,7
\bibitem[Salvaterra \& Burigana 2002]{Salvaterra2002} Salvaterra, R. \& Burigana, C. 2002,
\mnras, 336, 592
\bibitem[Sironi et al. 1990]{Sironi1990} Sironi, G., Limon, M.,
Marcellino, G., Bonelli, G., Bersanelli, M., Conti, G., Reif, K,
1990, \apj 357, ~301
\bibitem[Sironi et al. 1991]{Sironi1991} Sironi, G., Bonelli, G.,
Limon, M., 1991, \apj, ~378, 550
\bibitem[Sironi et al. 1999]{Sironi1999} Sironi, G., Boella, G.,
Bonelli, G., Gervasi, M., Vaccari, A. and Zannoni, M., 1999, AIP
Conf.Proc., ~476, 149
\bibitem[Smoot et al. 1985]{Smoot1985} Smoot, G.F., De Amici, G.,
Friedman, S.D. et al., 1985, \apj, ~291, L23
\bibitem[Stankevich et al. 1970]{Stankevich1970} Stankevich,
K.S., Wielebinski, R., Wilson, W.E., 1970, Australian J. Phys.,
~23, 529
\bibitem[Tartari et al. 2008]{Tartari2008} Tartari, A., Zannoni, M., Gervasi, M.,
Boella, G., Sironi, G., 2008 \apj, {\em submitted}
\bibitem[US Standard Atmosphere 1976]{US_STANDARD_ATMOSPHERE_1976}
US Standard Atmosphere. 1976, US GPO, 1976 O-599-256
\bibitem[Wolleben et al. 2006]{Wolleban2006} Wolleben, M., Landecker,
T. L., Reich, W., Wielebinski, R., 2006, A\&A, 448, 411
\bibitem[Zannoni et al. 1999]{Zannoni1999} Zannoni, M., Boella, G.,
Bonelli, G., Cavaliere, F., Gervasi, M., Lagostina, A., Passerini
A., Sironi, G. and Vaccari, A., 1999, AIP Cof. Proc., ~476, 165
\end{thebibliography}
\end{document}